\documentclass[%
 reprint,
 amsmath,amssymb,
 aps,
]{revtex4-2}

\usepackage[T2A]{fontenc}
\usepackage[utf8]{inputenc}

\usepackage{xcolor}
\usepackage{csquotes}
\usepackage{graphicx}% Include figure files
\usepackage{dcolumn}% Align table columns on decimal point
\usepackage{bm}% bold math
\graphicspath{{Figures/}}

\begin{document}

\preprint{APS/123-QED}

\title{Coherent Tunneling by Adiabatic Passage in Silicon Nitride based Integrated Waveguide Structures}

\author{Olga~Borovkova}
 \email{o.borovkova@rqc.ru.}
\affiliation{%
Russian Quantum center, Skolkovo, Moscow Region 143025, Russia%\\This line break forced% with \\
}%

\author{Junqiu~Liu}%

\affiliation{ 
International Quantum Academy, Shenzhen 518048, China%\\This line break forced with \textbackslash\textbackslash
}%
\affiliation{ 
Hefei National Laboratory, University of Science and Technology of China, Hefei 230088, China%\\This line break forced with \textbackslash\textbackslash
}
 
\author{Dmitry~Chermoshentsev}
 
\affiliation{%
Russian Quantum center, Skolkovo, Moscow Region 143025, Russia%\\This line break forced% with \\
}%
\affiliation{%
Moscow Institute of Physics and Technology, 141701, Dolgoprudny, Russia%\\This line break forced% with \\
}%

\author{Valery~Lobanov}
 
\affiliation{%
Russian Quantum center, Skolkovo, Moscow Region 143025, Russia%\\This line break forced% with \\
}%

\author{Igor~Bilenko}
 
\affiliation{%
Russian Quantum center, Skolkovo, Moscow Region 143025, Russia%\\This line break forced% with \\
}%
\affiliation{%
Faculty of Physics, Lomonosov Moscow State University, Leninskie Gory, Moscow 119991, Russia%\\This line break forced% with \\
}%


\begin{abstract}
Nowadays silicon nitride photonic integrated circuits serve as a mature platform for numerous applications. Planar waveguides and directional couplers made by CMOS-compatible technology are its basic elements. Here we demonstrate the possibilities of efficient light routing and transfer provided by the integrated planar Si$_3 $N$_4$ waveguides structure based on the coherent tunneling by adiabatic passage (CTAP) at the $1.55 \mu m$ telecom band. We addressed both high- and low-confinement silicon nitride CTAP structures and proved high efficiency of light routing in them. The mechanisms that limit the light control efficiency have been revealed. The accessible parameters of such structures have been determined. Besides that, there was proposed the original hybrid Si$_3 $N$_4$~- Si - Si$_3 $N$_4$ waveguides structure providing the enhanced efficiency and flexibility of the CTAP in comparison with the single-material Si$_3 $N$_4$ waveguide structures.
\end{abstract}

\maketitle

%\tableofcontents

\section{Introduction}

Coherent tunneling processes, which provide the coherent transport of population within an atomic or molecular system, received their initial development as the laser-based methods of efficient and selective transfer of population between quantum states \cite{Gaubatz:1990, Bergmann:1998, Shapiro:1988, Tannor:1986, Bergmann:2019, Lee:23}. In recent years they have found their analogies and applications in photonics, because they open up new opportunities to control the flow of light and are usually robust to imperfections and manufacturing perturbations that is appealing in integrated photonics. The stimulated Raman adiabatic passage (STIRAP) \cite{Gaubatz:1990, Bergmann:1998} is one of the known coherent tunneling processes, where the coupling between the initial and final quantum states occurs via an intermediate state.

The STIRAP has an optical analogy - coherent tunneling by adiabatic passage (CTAP) \cite{Longhi:2007, Longhi:2009}.  It represents the adiabatic evolution of light via the dark state of the three-state system. CTAP usually operates with a photonic transfer in coupled waveguides. In particular, a set of two weakly-curved and one straight coupled waveguides fulfill a perfect transfer of the light flow between the outer curved waveguides with negligible excitation of the intermediate straight one \cite{Longhi:2007, Lahini:2008}. It is reached by means of so-called counter-intuitive scheme, when coupling between the unexcited (or idle) waveguides appear before the excited waveguide becomes coupled with one of them. As a result, the transient waveguide is not excited via coupling, but just serve for light transfer. An advantage of such scheme is that it preserves high efficiency of the excitation transfer even for lossy intermediate waveguide. On the contrary, in the intuitive waveguide scheme the absorption properties of the intermediate waveguide become crucial because light propagates in it at a perceptible distance.

Different designs of integrated photonic devices resort to CTAP approach demonstrating polychromatic beam splitting \cite{Dreisow:2009}, spectral filters \cite{Menchon:2013}, high-density waveguides with suppressed cross-talk \cite{Mrejen:2015}, directional couplers \cite{Paspalakis:2006, Ramadan:1998}, etc. Besides, the optical analogy of STIRAP was demonstrated in nonlinear optical systems \cite{Lahini:2008, Wu:2014, Evangelakos:2023}.

CTAP has been demonstrated in both low- and high-contrast-refractive-index dielectric waveguides. The low-contrast-index waveguides satisfy adiabaticity conditions at a few millimeter-long distances \cite{Longhi:2007} that does not meet the compactness requirements for integrated photonic devices. For this purpose, the high-contrast-index waveguides are preferrable. Thus, in \cite{Mrejen:2015} the adiabatic protocol was realized in silicon waveguides. Ubiquitous use of silicon is dictated, first of all, by its leading role in microelectronics. However, this material exhibits strong two-photon absorption at the telecommunication frequencies, which makes the reseachers want to look for new materials for integrated photonics. A promising candidate for this role is a silicon nitride, Si$_3$N$_4$  \cite{Levy:2010, Moss:2013, Song:2016, Wu:2020, Liu:2021, Xiang:2022}. It is well-known for its ultralow losses in the spectral range from visible to mid-IR. Due to its high refractive index, Si$_3$N$_4$ waveguides support tight localization of light in silicon dioxide cladding \cite{Kondratiev:2023}. All this is accompanied by strong Kerr nonlinearity in comparison with SiO$_2$ material \cite{Spencer:2014}. 

To the best of our knowledge the CTAP have not been studied in the silicon nitride platform yet. This platform like other types of the photonic integrated circuits (PICs) requires the efficient junctions and interconnects. However, the planar geometry of PIC bans the channels intersections. The adiabatic passage of light without the excitation of the transient waveguide in the CTAP scheme allows one to create the coupling between the channels even though they are separated from each other.

Besides that, the CTAP concept is perspective for the effective coupling elements for microring resonators in silicon nitride PICs \cite{Moss:2013, Xiang:2022}. Also, it may be interesting for the photonic molecules and their dispersion properties control. Such systems of coupled microresonators serve as the efficient system for realization of different processes of nonlinear optics \cite{Rebolledo:23, Sanyal:2024}.

In this work, we consider the structure of silicon nitride waveguides buried into silicon dioxide substrate that adiabatically approach to each other providing the coherent tunneling of light. We consider two cases, high- and low-confinement waveguides, and analyze light dynamics and light transfer efficiency for both of them at telecom frequency of $1.55~\mu m$. Conditions for the efficient light routing are revealed. The efficiency of the CTAP scheme in the hybrid structure with the central waveguide made of another material is addressed as well. Such hybrid CTAP structure opens up the additional possibilities for the light transfer control as soon as the central waveguide can be made of the material with strong nonlinearity or significant magnetic response.

\section{Silicon Nitride Integrated CTAP Scheme}

\subsection{Design}

\begin{figure}[htbp]
\centering
\includegraphics[width=0.9\linewidth]{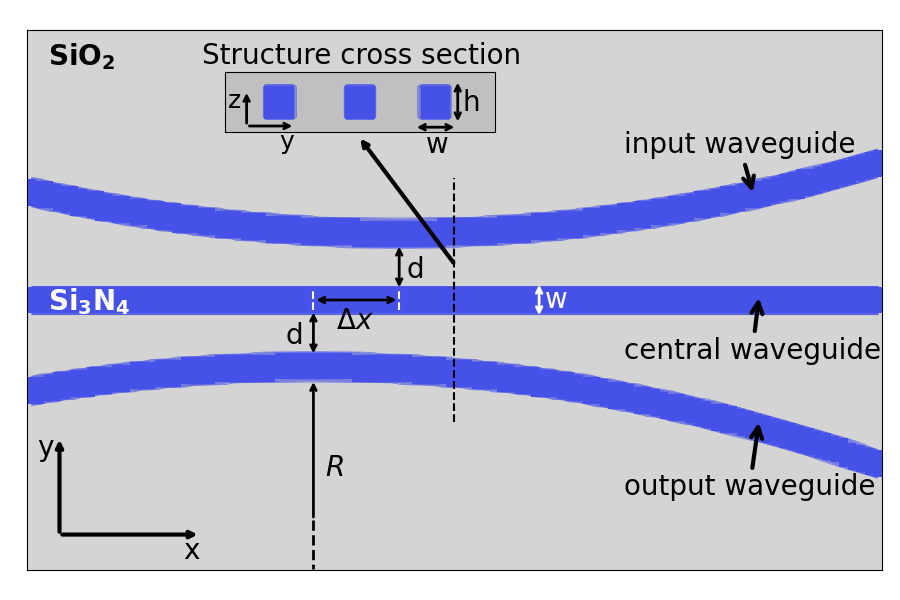}
\caption{Scheme of the addressed CTAP structure. Gray areas refer to the fused silica substrate, and blue color denote the silicon nitride waveguides. The inset shows the structure cross section. Horizontal and vertical scales are different.
%Folder: Fig.1
}
\label{fig:scheme}
\end{figure}

The principal scheme of the integrated Si$_3 $N$_4$ CTAP structure consisting of two bent (input and output) and one straight (transient) waveguides is given in Fig.~\ref{fig:scheme}. This plot contains the top view to the structure ($xy$ plane) and an inset where the cross section is given ($yz$ plane). Hosting silica \cite{Palik:1998} is shown by gray area, and the blue color indicates the waveguides made of silicon nitride \cite{Kischkat:2012}. The waveguides have square cross-sections (the parameters $w$ and $h$ refer to the width and the height of the waveguide, correspondingly) as it is shown in the inset in Fig.~\ref{fig:scheme}. The dashed line cutting the main figure indicates where the cross section is taken. 

Light propagates through the triple-waveguide structure from left to the right along $x$-axis. Two bent waveguides consequently approach the straight waveguide in the center. They can be parts of the integrated microrings as well. The curvature radius, $R$, is similar for both bent waveguides in Fig.~\ref{fig:scheme}, but can be taken different. To simplify our explanations let's denote two outer bent waveguides as  \enquote{input waveguide} and \enquote{output waveguide}. At the beginning the input waveguide is excited. The flow of light inside it propagates along the waveguide, and due to its curvature slowly approaches the central straight waveguide, so-called \enquote{central waveguide}. The central and output waveguides are idle at the beginning, and become excited just due to the light transfer from neighbor waveguides. It is employed the so-called counter-intuitive scheme, where the unexcited output waveguide firstly approaches the unexcited central one so the coupling between them appears, and then the excited input waveguide comes close to the central waveguide. Parameter $d$ in Fig.~\ref{fig:scheme} is the minimal distance between two neighbor waveguides. Short white dashed lines cutting the central waveguides indicate the distances along $x$-axis where the waveguides are in the most proximity. The distance between these points is denoted as $\Delta x$. 

We perform the numerical simulation of the waveguide modes propagation in the structure by means of the bidirectional eigenmode propagation method. Varying the values of parameters $R, d, w, h, \Delta x$ it is possible to determine the optimal set of parameters by minimizing the light intensity at the outcome of the input and central waveguides under the condition of a minimum loss in the entire structure. In our numerical simulations we took the parameters of the real silicon nitride waveguide structures \cite{Xiang:2022, Blumenthal:2018, Pfeiffer:16, Mumlyakov:2024}. The bent waveguides in the CTAP scheme could be the parts of the ring waveguides or microresonators \cite{Rebolledo:23, Sanyal:2024}.

\begin{figure}[htbp]
\centering
\includegraphics[width=\linewidth]{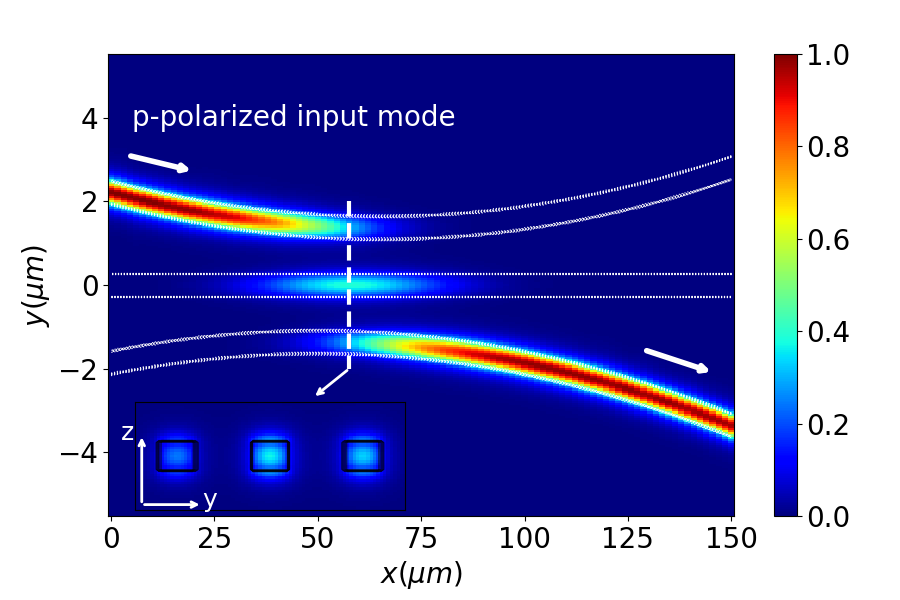}
\caption{Intensity $(|E|^2)$ distribution in the silicon nitride waveguides for $p$-polarized input waveguide mode at the wavelength of $1.55\mu m$. White dotted lines indicate the contours of Si$_3$N$_4$ waveguides. Horizontal and vertical scales are different. White arrows denote the light propagation direction. The inset shows the light intensity distribution in the structure cross section according to the white dashed line. The colormap in the inset is the same as in the main plot.
%Folder: Fig.2
}
\label{fig:OptParam}
\end{figure}

In Fig.~\ref{fig:OptParam} an example of the intensity distribution of $p-$polarized mode of the silicon nitride waveguide at $1.55\mu m$ in the CTAP setup with optimized parameters is given. White dotted lines indicate the contours of the Si$_3$N$_4$ waveguides in SiO$_2$ substrate. Here waveguide width is $w=0.55\mu m$, its thickness is $h=0.55\mu m$ as well, the minimal distance between neighboring waveguides is $d=0.815\mu m$, radius of curvature $R=2.5mm$, and $\Delta x=15.2\mu m$. Note that the length required for the light transfer for the parameters indicated above is about $110 \mu m$ that is much smaller than in case of low-contrast-refractive index case \cite{Longhi:2007}. For such set of waveguide parameters the $p-$polarized mode of the waveguide has $n_{eff}=1.5647$. The setting is optimized for the $p-$polarized input waveguide mode and operating wavelength of $1.55\mu m$. Here and below, we explore the $p-$polarized input mode alone due to the fact that the transmittance properties of the scheme under consideration weakly depends on the polarization state of light.

The values of the width, height, and gap width in Fig.~\ref{fig:OptParam} are applicable for the manufacturing. However, the high-confinement waveguides usually operate with the waveguide thickness of about $0.8-1.0 \mu m$ (see, for instance, \cite{Xiang:2022, Blumenthal:2018, Pfeiffer:16, Mumlyakov:2024}). But with the growth of the thickness of Si$_3$N$_4$ layer, the width of the waveguides should be increased as well to avoid the defects of sidewalls, an appearance of the unwanted deviations from the perfectly vertical sidewalls and etc. However, for adiabatic process considered in this work it means also that the minimal gap width should be dramatically diminished, because the coupling coefficient decreases with the growth of the waveguide thickness. For instance, the waveguide of square profile with $w=1.0 \mu m$ and $h=1.0 \mu m$ requires the gap distance as small as $0.265 \mu m$. Such configuration is hard or almost impossible for fabrication. 

\begin{figure}[htbp]
\centering
\includegraphics[width=0.7\linewidth]{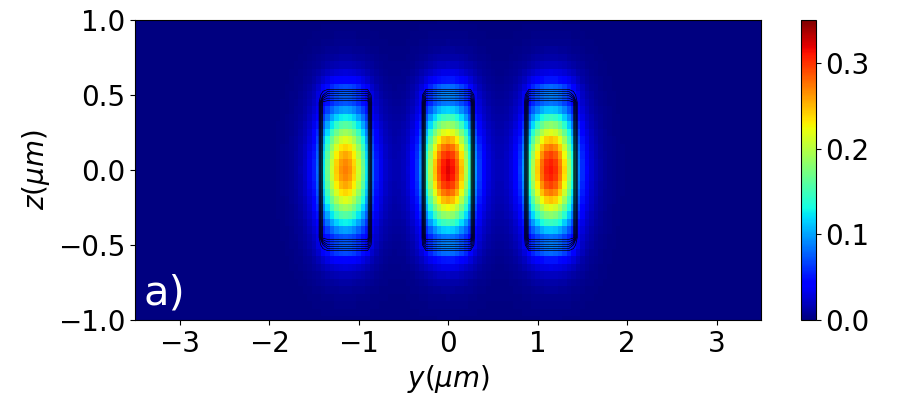}
\includegraphics[width=0.7\linewidth]{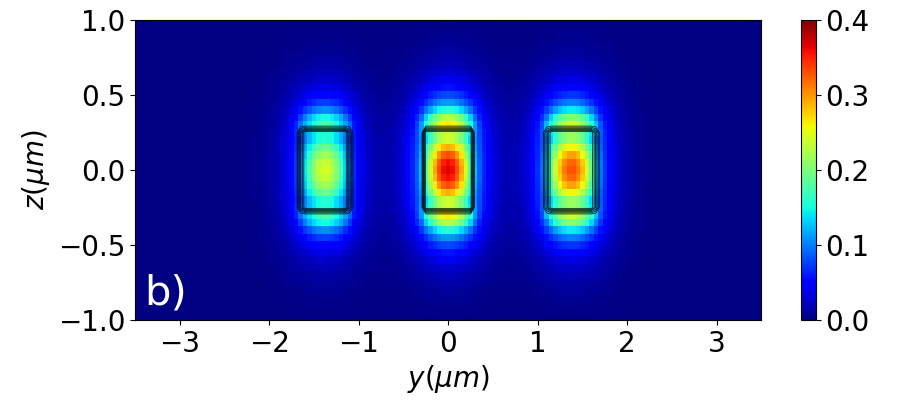}
\includegraphics[width=0.7\linewidth]{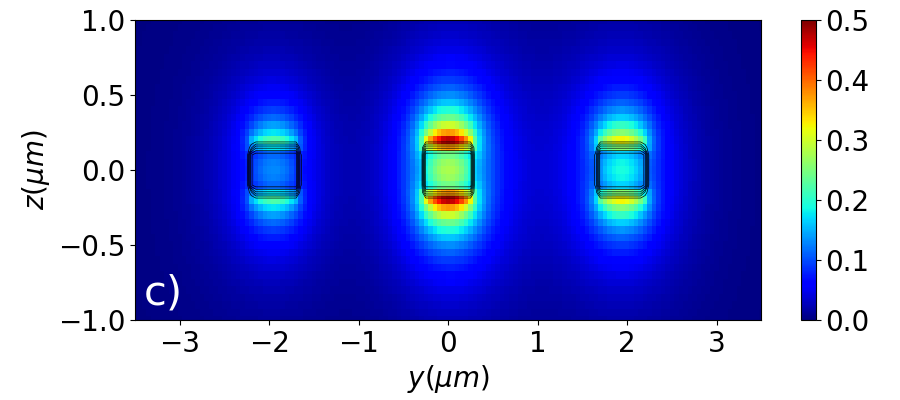}
\caption{Intensity $(|E|^2)$ distribution in the cross-section of the silicon nitride waveguides with the width $w=0.55 \mu m$ and height a) $h=1.0 \mu m$, b) $h=0.55 \mu m$ and c) $h=0.3 \mu m$ at $x=57.6 \mu m$ (corresponds to the cross-section in the Fig.~~\ref{fig:OptParam}) for $p$-polarized input waveguide mode at the wavelength of $1.55\mu m$. Black lines indicate the contours of Si$_3$N$_4$ waveguides in the fused silica. 
%Folder: Fig.2CS
}
\label{fig:CrosSec}
\end{figure}

In Fig.~\ref{fig:CrosSec} the light distribution in the cross sections of the CTAP structure with three different thicknesses are presented. These plots allow one to trace the transfer from high-confinement modes (Fig.~\ref{fig:CrosSec}a) to low-confinement ones (Fig.~\ref{fig:CrosSec}c). In all these cases the width of the waveguides is the same, namely, $w=0.55 \mu m$, but the thickness, $h$, and the gap width, $d$, are chosen so to support the CTAP light transfer with maximum efficiency for different values of $h$. All the cross sections are taken at $x=57.6 \mu m$, exactly in the middle between the coordinates where the lateral waveguides approaches to the central one. One can see that in Fig.~\ref{fig:CrosSec}a the modes are strongly localized and the light propagates just inside the waveguides. At the same time, in Fig.~\ref{fig:CrosSec}c the waveguides serve just for the guiding of light residing mostly in the surrounding SiO$_2$ medium. Interestingly, this not only does not prevent the coherent transfer of light in the low-confinement waveguides, but also allows to fabricate them with the wider gap between them that is more convenient for the manufacture.

Thus, it is interesting to explore what are the possible and optimal parameters of the waveguide and the gap between them for the effective CTAP and whether it is possible to focus on the low-confinement waveguides to realize the adiabatic light tunneling.

\subsection{Optimal Parameters of Silicon Nitride CTAP Structure}

\begin{figure}[htbp]
\centering
\includegraphics[width=\linewidth]{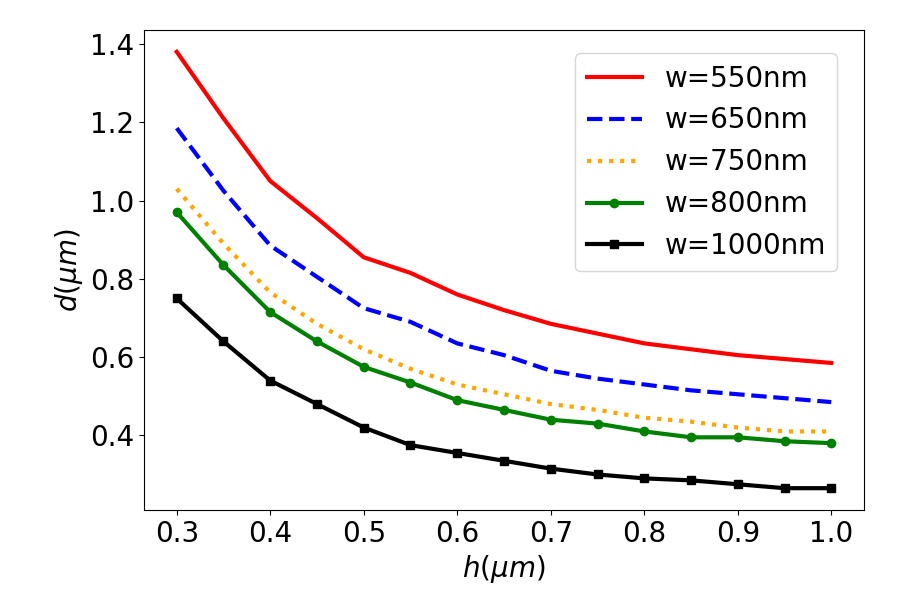}
\caption{Dependence of minimal gap, $d$, versus the thickness of the waveguides, $h$ providing the maximum efficiency of the CTAP for different values of waveguide width, $w$. The dependence was obtained for $p$-polarized input waveguide mode at the wavelength of $1.55\mu m$.
%Folder: Fig.2n
}
\label{fig:dvsh}
\end{figure}

The low-confinement of light in thin waveguides allows to increase the quality factor of silicon nitride microresonators, since a significant part of the field resides in the surrounding environment with lower losses \cite{Jin:2021}. One might expect that this should provide increased efficiency of light tunneling through the adiabatic process due to higher coupling. So, it is interesting to explore the properties of the CTAP in low-confinement silicon nitride waveguides and compare the results with the high-confinement case. For this purpose, we fix the radius of curvature of lateral waveguides $R=2.5mm$ and the spatial shift between them along the propagation direction is $\Delta x=15.2\mu m$, all other parameters have been varied (see Fig.~\ref{fig:scheme}). For different values of waveguide width, $w$, we found the pair (waveguide thickness, minimal gap) that provides the maximum efficiency of the light transfer between lateral waveguides. The result for five selected values of $w$ is presented in Fig.~\ref{fig:dvsh}. Note that other values of waveguide width provide qualitatively the same dependencies.

In Fig.~\ref{fig:dvsh} there are the dependencies of the parameters providing maximum efficiency of the CTAP for the structure with waveguide width $w=0.55~\mu m,~0.65~\mu m,~ 0.75~\mu m,~ 0.8~\mu m$ and $1.0~\mu m$. The thickness of the silicon nitride waveguides ranged from $0.3~\mu m$ to $1.0~\mu m$. At the operating wavelength $1.55~\mu m$ and in the addressed range of $w$ values, the waveguide of thickness $h=1.0~\mu m$ supports the highly localized mode (see Fig.~\ref{fig:CrosSec}a). Meanwhile, the thickness of $h=0.3~\mu m$ refers to the low-confinement of the mode if other parameters remain similar (see Fig.~\ref{fig:CrosSec}c). 

It is clear from Fig.~\ref{fig:dvsh} that for all values of the waveguide width the minimal gap between the neighboring waveguides, $d$, increases with decreasing waveguide thickness, $h$. Talking about the black curve ($w=1.0~\mu m$) one can see that in case of waveguide thickness of $h=1.0~\mu m$ the minimum required gap is $d=0.265~\mu m$, which is almost impossible for the fabrication with satisfied quality of the waveguides. At the same time, for the thickness less than $0.520\mu m$ the required gap for efficient light transfer exceeds $d=0.4\mu m$, which is appropriate for the fabrication. Even more suitable is the pair of parameters $h=0.3\mu m$ and $d=0.75\mu m$ at the left tail of the considered curve which supports the weakly localized mode. 

\begin{figure}[htbp]
\centering
\includegraphics[width=\linewidth]{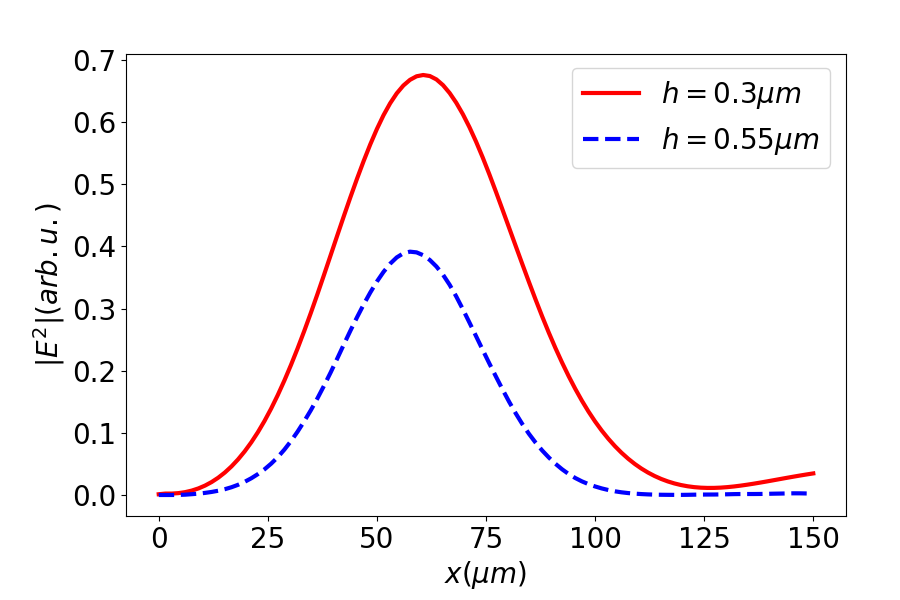}
\caption{Intensity $|E|^2$ at the center of the middle waveguide ($y=0~\mu m$, $z=0~\mu m$) versus the longitudinal coordinate $x$ in the low-confinement (solid red line) and high-confinement (dashed blue line) silicon nitride waveguides for $p$-polarized input waveguide mode at the wavelength of $1.55~\mu m$. Other parameters are $w=0.55~\mu m$, $d=1.38~\mu m$, and $R=2.5~mm$.
%Folder: Fig.3lc
}
\label{fig:OptParamLowConf}
\end{figure}

However, in case of low-confinement the process of the waveguide coupling after the adiabatic tunneling should be taken into account as well. As a result, some energy from the output waveguide transfer back to the central and even input wavegide. In Fig.~\ref{fig:OptParamLowConf} there is a dependence of the intensity $|E|^2$ at the central axis of the middle waveguide versus the longitudinal coordinate $x$ for two cases:  red solid line refers to low-confinement case ($w=0.55 \mu m$, $h=0.3 \mu m$ and $d=1.38 \mu m$) and the blue dashed line corresponds to high-confinement case (as in Fig.~\ref{fig:OptParam}). One can see that despite the huge gap between the neighbor waveguides the light re-couple to the central due to the strong evanescent field in the cladding of the waveguides and after $x=125\mu m$ we observe the gradual increase of the intensity in the central waveguide.

Another exciting aspect is how the coupling length differs for low- and high-confinement CTAP schemes. Coupling length parameter ($L_c$) could be estimated, as  it is usually used in two-waveguide cases, as a double distance, where the intensity in the input waveguide decreases by 2 times. For the structure under consideration the parameter $L_c$ is greater for low-confinement case than for the high-confinement one. For instance, it reaches $82 \mu m$ for the waveguides with $h=0.3 \mu m$ (for waveguide width $w=0.55\mu m$) and then gradually decreases with the growth of the waveguide height. The corresponding high-confinement case with $h=1.0 \mu m$ and $w=0.55\mu m$ shows the intensity decrease by 2 times at the distance of just $L_c=70 \mu m$.

However, as soon as we analyze the three-waveguide structure the approach given above can miss some light transfer processes in the central and output waveguides. For this purpose, we also estimated the distance along $x$-axis, where the intensity magnitude changes in any of three waveguides. In other words, the later parameter shows the maximum distance at which the modes in waveguides \enquote{feel} the electromagnetic field in other waveguides. This approach gives the coupling length of $116 \mu m$ for the waveguides with $h=0.3 \mu m$ and $89 \mu m$ for the waveguides with $h=1.0 \mu m$ for the waveguides with $w=0.55\mu m$. 

Thus, although the decrease of the thickness of the guiding channels results in the increase of the possible minimal gap between the waveguides, it is accompanied by the parasitic re-coupling lowering the efficiency of the whole process and by the growth of the coupling length in the structure. This trade-off should be taken into account when the adiabatic tunneling is used in the waveguide scheme.

\subsection{CTAP in the Non-Symmetric Structures}

Due to the adiabaticity of the process under consideration it admits the symmetry breaking of the structure and can be realized even if the structure is slightly non-symmetric. We explored the tolerance of the CTAP to the difference of the width and radius of the input and output waveguides.

\begin{figure}[htbp]
\centering
\includegraphics[width=\linewidth]{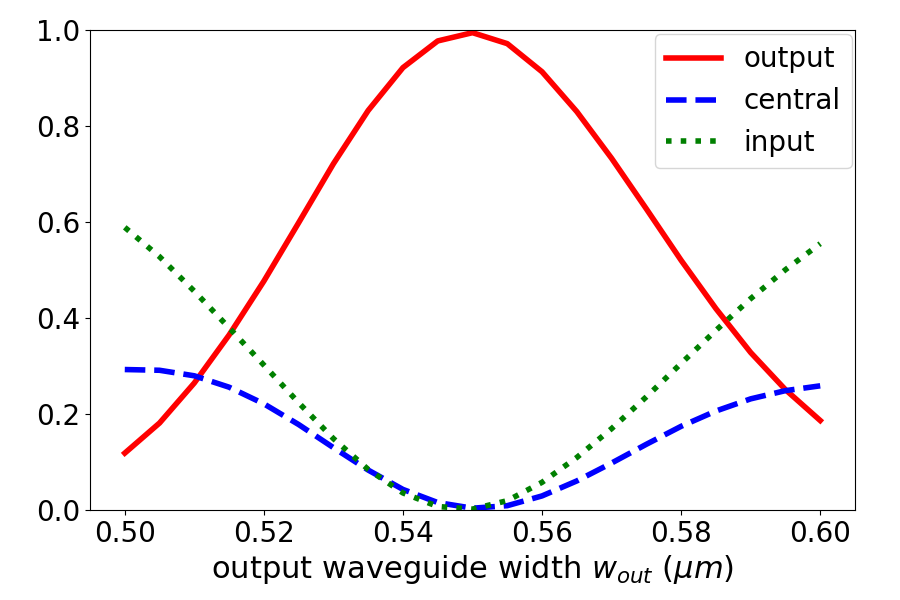}
\caption{The intensity fraction at the output of each waveguide with respect to the width of the output waveguide $w_{out}$, when the input and central waveguides width is $w=0.55\mu m$, $h=1 \mu m$, $\Delta x=15.2\mu m$ and $R=2.5 mm$.
%Folder: Response_PRAppl
}
\label{fig:dwyOut}
\end{figure}

In Fig.~\ref{fig:dwyOut} we presented how the output intensity redistributes between the waveguides if the width of the output waveguide width is varied. As soon as the parameters were selected to provide maximum light transfer for $w=0.55 \mu m$ we observe there a maximum of red line. However, one can see that for $w$ in the range of $0.54 \mu m - 0.56 \mu m$ the output waveguide will account for over $90\%$ of the power, and $w$ from $0.53 \mu m$ to $0.57 \mu m$ provides over $80\%$ of the power in the output waveguide.

\begin{figure}[htbp]
\centering
\includegraphics[width=\linewidth]{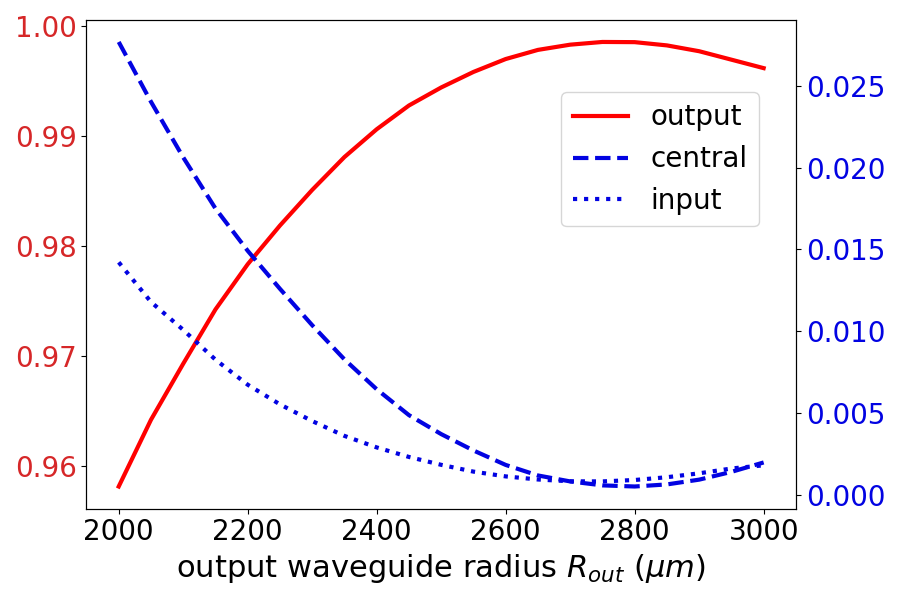}
\caption{The intensity fraction at the output of each waveguide with respect to the radius of curvature of the output waveguide, $R_{out}$, when the input waveguide radius is $R=2.5 mm$, $w=0.55\mu m$, $h=1 \mu m$, and $\Delta x=15.2\mu m$.
%Folder: Response_PRAppl
}
\label{fig:dROut}
\end{figure}

Even more tolerant the CTAP is to the asymmetry of the bent waveguides radii of curvature. In Fig.~\ref{fig:dROut} we show the output intensity in the waveguides when we vary the radius of curvature of the output waveguide from $2.0 mm$ to $3.0 mm$. The increase of the radius of curvature doesn't affect the light routing at all, and the decrease of $R$ of the output waveguide up to $2.0 mm$ results in the decrease of the intensity in the output waveguide by $4\%$. 

Therefore, the CTAP allows to perform the coupling and light routing between elements with different parameters.

\section{Hybrid Silicon Nitride - Silicon Integrated CTAP Scheme}

Besides the re-coupling in the low-confinement regime the light transfer shown above is accompanied by the small, but still non-zero excitation of the central waveguide (see Figs.~\ref{fig:OptParam},~\ref{fig:CrosSec},~\ref{fig:OptParamLowConf}) despite the application of the counter-intuitive CTAP scheme. To overcome this obstacle we consider the hybrid silicon nitride - silicon integrated CTAP scheme where the central waveguide is substituted by the waveguide made of silicon. Silicon is a semiconductor and changing the carrier density by means of PN junctions it is possible to realize a high speed optical modulation \cite{Malacarne:2014}. From the other hand, silicon exhibits strong non-linear two-photon absorption in the $1.55 \mu m$ telecom band, hence adiabatic tunneling allows to avoid additional losses. 

However, as soon as the silicon has higher refractive index than the silicon nitride, the central waveguide should have smaller width to provide the light tunneling between the lateral waveguides. We use the width of the central silicon waveguide, $w_{Si}$, as a variable and explore the character and efficiency of the light transfer with respect on it.

\begin{figure}[htbp]
\centering
\includegraphics[width=\linewidth]{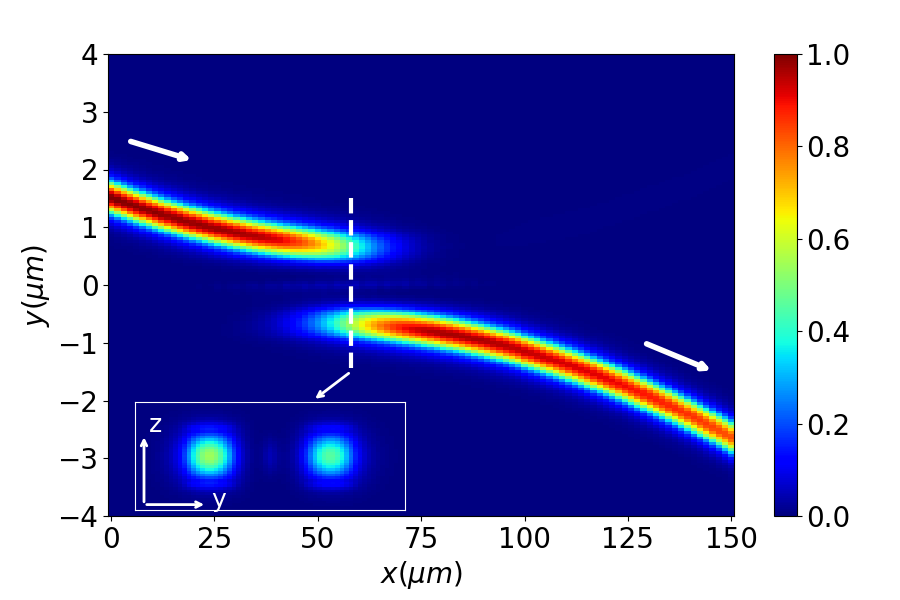}
\caption{High-confinement intensity $(|E|^2)$ distribution in the hybrid CTAP structure of silicon nitride lateral waveguides and silicon central waveguide for $p$-polarized input waveguide mode at the wavelength of $1.55\mu m$. Horizontal and vertical scales are different. White arrows denote the light propagation direction. The inset shows the light intensity distribution in the structure cross section according to the white dashed line. The colormap in the inset is the same as in the main plot.}
%Folder: Fig.6Si
\label{fig:SiHighConf}
\end{figure}

Let's consider the hybrid CTAP scheme, where the lateral waveguides are made of silicon nitride and have fixed width of $w=0.55 \mu m$. The thickness of either lateral or central waveguides are the same. First of all, let's address the high-confinement waveguides with the thickness $h=1.0 \mu m$. In Fig.~\ref{fig:SiHighConf} it is shown the light dynamics in the corresponding waveguide setting. The width of the central waveguide is $w_{Si}=0.1 \mu m$, the minimal gap between the neighbor waveguides is $d=0.32 \mu m$. Also, for the maximum efficiency of the light transfer the shift of the lateral waveguides is $\Delta x = 16 \mu m$ (see also Fig.~\ref{fig:scheme}). One can see that the light intensity in the central waveguide is almost negligible, and the majority of light is concentrated in the lateral ones. As it is clear from the inset with the light distribution in the cross section of the structure, the light is highly confined in the waveguides.

On the one hand, the higher refractive index in the central waveguide requires to diminish the width of the central waveguide, as it happens in the considered case from $0.55 \mu m$ to  $0.1 \mu m$. Meantime, the optical and dielectric properties  of the central waveguide are still important for the light transfer process. Manipulating them by means of the external magnetic field or by excitation the mode at the different wavelength in the central waveguide one can control the light routing in the whole integrated structure, e.g. tunneling efficiency.  On the other hand, as one can see in the inset in Fig.~\ref{fig:SiHighConf} the light is concentrated in the lateral waveguides. So, the hybrid structure prevents light from the penetration into the central waveguide and minimize the light intensity in it in contrast to purely silicon nitride structure addressed above.

\begin{figure}[htbp]
\centering
\includegraphics[width=\linewidth]{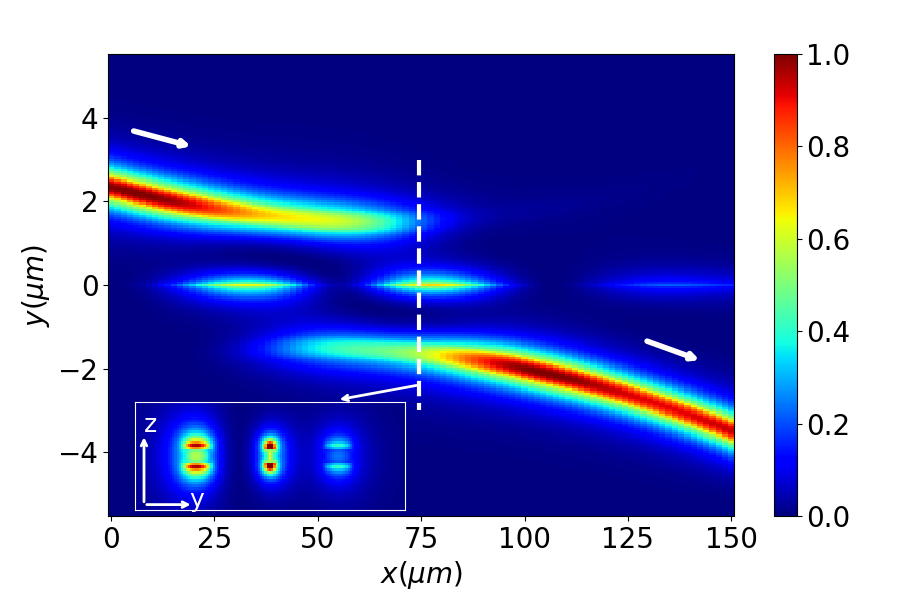}
\caption{Low-confinement intensity $(|E|^2)$ distribution in the hybrid CTAP structure of silicon nitride lateral waveguides and silicon central waveguide for $p$-polarized input waveguide mode at the wavelength of $1.55\mu m$. Horizontal and vertical scales are different. White arrows denote the light propagation direction. The inset shows the light intensity distribution in the structure cross section according to the white dashed line. The colormap in the inset is the same as in the main plot.}
%Folder: Fig.7Si
\label{fig:SiLowConf}
\end{figure}

The low-confinement guiding is also possible in the hybrid CTAP structure. In Fig.~\ref{fig:SiLowConf} there is an example of the light transfer in the waveguide structure with the thickness $h=0.3 \mu m$, $w=0.55 \mu m$, $w_{Si}=0.1 \mu m$, minimal gap width $d=1.17 \mu m$, and spacing $\Delta x=15.2 \mu m$. In the inset it is shown the light intensity distribution in the cross-section of the waveguide structure. One can see that the light is mostly localized in the nesting material below and above the waveguides. Due to this fact the dynamics of light propagation in the low-confinement CTAP structure is different from the process given above in Fig.~\ref{fig:SiHighConf}. Namely, we observe several peaks of intensity in the central waveguide that happens due to the re-coupling process, because the modes reside outside the waveguide and they are not limited by the potential barrier. Therefore, the coherent adiabatic light tunneling in the low-confinement waveguides is inevitably accompanied by the simultaneous coupling process that change the light dynamics.

%\textcolor{blue}{Посчитать и привести значение длины взаимодействия.}

\section{Conclusion}

We address the coherent tunneling by adiabatic passage in waveguides made of silicon nitride, a material that is actively used in integrated photonics nowadays. The efficiency of the CTAP in Si$_3$N$_4$ waveguides in the spectral range of telecommunication frequency is demonstrated in both regimes of high- and low confinement for the three-waveguide system. It was found that for the particular value of the waveguide width there is the optimal combination of waveguide thickness and minimal gap that provides the maximum efficiency of the light transfer between lateral waveguides. The minimal gap between the neighboring waveguides increases with decreasing of the waveguide thickness. It was found that the coupling length decreases with the growth of the light confinement in the waveguides and the high-confinement case is better from the compactness point of view. Also, it was revealed that high-confinement regime requires tightly packed waveguides that is a challenge for the fabrication. From the other hand, the low-confinement regime suffers from parasitic excitation of the central waveguide by re-coupling effect. So, the intermediate case may be preferable for a wide range of applications. 

It was revealed that the CTAP process is tolerant to the asymmetry of bent waveguides. It was found that the decrease of the output waveguide radius of curvature from $2.5 mm$ to $2.0 mm$ results in decrease of the intensity in that waveguide by only $4\%$. On the same time the increase of such waveguide radius of curvature has no effect at all. The difference in the width of the lateral waveguides is more critical, but it also admits the range of $20 nm$ when the intensity in the output waveguide decreases by less than $10\%$. 

The coherent adiabatic tunneling can be implemented in the integrated circuits to realize the efficient junctions and interconnects of different waveguides even though these channels are separated from each other.

Besides that, the original hybrid silicon nitride - silicon~- silicon nitride CTAP structure has been proposed and studied in detail. It was found that this configuration is even more efficient because the excitation of the central waveguide is almost absent, allowing the avoidance of a two-photon absorption in silicon. The concept of the CTAP scheme with the central waveguide made of different material (or hybrid CTAP scheme) provides ample possibilities for the light routing control in the integrated circuits as soon as the central waveguide can be made of the material with strong nonlinearity or significant magnetic response. This approach can be used to create a wide range of different light switches.

The lateral waveguides could be the parts of the microrings whose coupling can be effectively modified at will by the transient waveguide. Such dynamic control of the coupling level between microrings will lead to dynamic control of the dispersion characteristics of photonic molecules that is important for the efficient realization of different processes of nonlinear optics like generation of solitons and optical frequency combs. 

\begin{acknowledgments}
This study was supported by Russian Science Foundation (project no. 24-22-00190). J. Liu acknowledges support from the National Natural Science Foundation of China (Grant No.~12261131503).
\end{acknowledgments}

\nocite{*}

\bibliography{apssamp}% Produces the bibliography via BibTeX.

\end{document}